\documentclass[twocolumn]{aastex701}
\usepackage{tabularx}
\usepackage{amsmath}
\usepackage{amssymb}
\usepackage{longtable}
\usepackage{multirow}
\usepackage{bm}
\usepackage{mhchem}
\usepackage{rotating}
\usepackage{xcolor}
\usepackage{float} 

\begin{document}

\title{A Climate-Constrained Bayesian Inverse Method for JWST Rocky Exoplanet Eclipse Spectra: A Case Study of LTT 1445A b}

\author[0000-0002-0413-3308]{Nicholas F. Wogan}
\affiliation{SETI Institute, Mountain View, CA 94043}
\affiliation{NASA Ames Research Center, Moffett Field, CA 94035}
\email{nicholas.f.wogan@nasa.gov}

\author[0000-0003-1240-6844]{Natasha E. Batalha}
\affiliation{NASA Ames Research Center, Moffett Field, CA 94035}
\email{natasha.e.batalha@nasa.gov}

\author[0000-0003-2775-653X]{Jegug Ih}
\affiliation{Space Telescope Science Institute, Baltimore, MD 21218}
\email{jih@stsci.edu}

\author[0000-0002-0746-1980]{Jacob Lustig-Yaeger}
\affiliation{Johns Hopkins APL, 11100 Johns Hopkins Rd, Laurel, MD 20723, USA}
\email{jacob.lustig-yaeger@jhuapl.edu}

\author[0000-0002-7352-7941]{Kevin B. Stevenson}
\affiliation{Johns Hopkins APL, 11100 Johns Hopkins Rd, Laurel, MD 20723, USA}
\email{kevin.stevenson@jhuapl.edu}

\begin{abstract}
  Determining whether temperate rocky exoplanets orbiting M stars retain atmospheres is currently a central goal of exoplanet astronomy. To this end, the James Webb Space Telescope has begun searching for atmospheres on these worlds with MIRI secondary eclipse spectroscopy and photometry. Here, we develop a novel climate-constrained Bayesian inference framework that yields atmospheric pressure and composition constraints from these datasets, while accounting for planetary, stellar, and model uncertainties. Our approach fits observations with model spectra derived from self-consistent pressure-temperature profiles at radiative-convective equilibrium, thus maximizing the information extracted from the data and providing more robust inferences than retrievals that use parameterized pressure-temperature profiles. We demonstrate the framework on the existing MIRI LRS eclipse spectrum of LTT 1445A b (1.34 $R_\oplus$ and $T_{\mathrm{eq}} \approx 431$ K). An atmosphere does not need to be invoked to explain the data, meaning a bare rock model produces an adequate fit. If the planet has an atmosphere, the $2\sigma$ upper limits on surface partial pressures are $\lesssim 1$ bar for an optically thin gas like O$_2$, N$_2$ or CO, $\lesssim0.1$ bar for CO$_2$, $\lesssim 10^{-3}$ bar for H$_2$O, and $\lesssim 10^{-4}$ bar for SO$_2$. Scheduled MIRI F1500W observations could detect one of the thicker atmospheres permitted by the existing data (1 bar O$_2$ and 0.01 bar CO$_2$), if a precision of 20 ppm or better is achieved. This case study demonstrates that climate-constrained Bayesian inversion can turn rocky-planet eclipse spectra into the quantitative constraints necessary to test population-level atmospheric retention hypothesis, like the cosmic shoreline.
\end{abstract}

\section{Introduction}

Do temperate rocky planets orbiting M dwarfs possess atmospheres? This is one of the most profound unsolved questions in planetary astronomy. M stars experience an extended pre-main sequence phase that irradiates planets with significant bolometric and X-ray/extreme ultraviolet (XUV) radiation, potentially completely eroding atmospheres \citep[e.g.,][]{KrissansenTotton2022}. Also, M dwarf systems host the only temperate rocky worlds amenable to atmospheric characterization with current facilities, like the James Webb Space Telescope (JWST), and just a handful of these promising targets are in the habitable zone. Our ability to study the composition of terrestrial exoplanet atmospheres or search for biosignatures in the near term hinges upon whether rocky planets orbiting M stars can hold onto volatile envelopes.

Through the 500-hour Rocky Worlds Director's Discretionary Time (DDT) program and numerous General Observers (GO) programs, JWST has been collecting data in search for rocky planet atmospheres. In particular, JWST has measured the dayside thermal emission of terrestrial exoplanets during secondary eclipse events with MIRI LRS and various photometry filters (${\sim}15$ $\mu$m). Bare, atmosphere-free planets emit strongly because they instantly re-radiate absorbed starlight, whereas planets with atmospheres can have reduced thermal emission when winds redistribute heat from the day to night side of the tidally locked planets \citep{Koll2019}. Furthermore, atmospheres can have unique molecular absorption (e.g., 15 $\mu$m CO$_2$ band) that can not be emulated by plausible bare rock surface compositions \citep{Paragas2025}.

Thus far, there are nine relatively temperate planets ($T_{\mathrm{eq},0} < 700$ K) with published JWST MIRI secondary eclipse observations. The following seven targets appear to have large eclipse depths interpreted as broadly consistent with bare, airless worlds: TRAPPIST-1 b \citep{Greene2023,Ih2023,Maurel2025}, TRAPPIST-1 c \citep{Zieba2023,Lincowski2023}, LHS 1140 c \citep{Fortune2025}, TOI-1468 b \citep{MeierValdes2025}, GJ 3929 b \citep{Xue2025}, GJ 486 b \citep{Mansfield2024}, GJ 1132 b \citep{Xue2024}. On the other hand, analyses of LHS 1478 b and LTT 1445A b indicate that the existing MIRI data do not necessarily rule out multi-bar atmospheres \citep{Wachiraphan2025,August2025}. In a far different planetary and stellar regime, MIRI observations have yielded tentative evidence for atmospheres on ultra-hot lava worlds ($T_\mathrm{irr} \gtrsim 1800$ K) orbiting G and K stars \cite[e.g.,][]{Coy2026,Teske2025}.

Most published MIRI observations of temperate rocky planets have been evaluated with forward-model comparisons targeted to the data. In this approach, studies typically use a 1.5-D climate model, such as \texttt{HELIOS} \citep{Malik2019}, to simulate the temperature profiles and corresponding emission spectra of a limited set of atmospheric compositions and surface pressures, often emphasizing \ce{CO2}- and \ce{H2O}-rich cases. The resulting model spectra are then compared to the data, usually with a $\chi^2$ statistic, to identify atmospheric scenarios that are consistent or not with the observations.

A forward-modeling analysis is valuable for building intuition, but it does not systematically marginalize over all parameters. In particular, sparse forward-modeling can miss viable regions of atmospheric composition and pressure space, and fixed assumptions about stellar properties, planetary properties, surface albedo, or heat redistribution can under characterize the uncertainty and impact atmospheric interpretation \citep{Fauchez2025,Hammond2025}. These limitations are especially important if MIRI eclipse observations are to be used to derive surface-pressure constraints across all M-dwarf rocky exoplanets observed by JWST to test population-level hypotheses, like the cosmic shoreline \citep{Zahnle2017,BertaThompson2025}.

A standard free-retrieval framework is not a clear solution to this problem. Although retrieval codes such as \texttt{POSEIDON} \citep{MacDonald2017,MacDonald2023} can explore large parameter spaces, they often treat the temperature-pressure (T-P) profile and atmospheric composition independently, without enforcing radiative-convective equilibrium and self-consistent heat redistribution. When the T-P profile and composition are decoupled in this way, current JWST rocky-planet eclipse spectra may not contain enough information to constrain surface pressure.

Here, to move beyond both targeted forward-model comparisons and standard free retrievals, we introduce a climate-constrained Bayesian inversion framework for interpreting JWST rocky exoplanet thermal emission spectra. Our approach first uses a 1.5-D climate code to compute a grid of self-consistent T-P profiles and corresponding emission spectra spanning a wide range of surface pressures, compositions, and other parameters. We then perform a grid-based Bayesian inversion, incorporating both model and planet/star uncertainties to derive quantitative constraints on the atmosphere. The framework also fits spectra with a bare-rock model, and uses model comparison to determine if the data prefers the more sophisticated atmospheric model.

We choose to demonstrate this climate-constrained inversion framework on the MIRI/LRS eclipse spectrum of LTT 1445A b published by \citet{Wachiraphan2025}. This planet is an ideal case study for two reasons. First, \citet{Wachiraphan2025} argued that more detailed atmospheric modeling will likely be required to characterize any atmosphere that might remain on LTT 1445A b. Their forward-modeling analysis that used a climate code was limited to pure \ce{CO2} atmospheres. Second, compared to other JWST rocky targets, LTT 1445A b has relatively high surface gravity and low estimated cumulative XUV irradiation, making it a potentially favorable candidate for atmospheric retention when considering the XUV cosmic shoreline \citep{Zahnle2017}. Thus, LTT 1445A b provides both a practical demonstration of our method and a scientifically important test case for atmospheric retention around M dwarfs.

\section{Methods} \label{sec:methods}

\subsection{Overview of the inference framework}

We fit two competing models to the MIRI LRS eclipse spectrum of LTT 1445A b: one in which the planet has an atmosphere, and one in which it is a bare rock. Section \ref{sec:methods_atmos} describes the atmospheric model, Section \ref{sec:methods_bare} describes the bare-rock model, and Section \ref{sec:methods_inversion} details the Bayesian inversion. The posterior from the atmospheric fit provides constraints on surface pressure, atmospheric composition, and other model parameters. We also compare the Bayesian evidences of the two models to assess whether or not the data favor the presence of an atmosphere. 

\subsection{Atmospheric model} \label{sec:methods_atmos}

\subsubsection{Model overview}

\begin{figure*}
  \centering
  \includegraphics[width=0.9\textwidth]{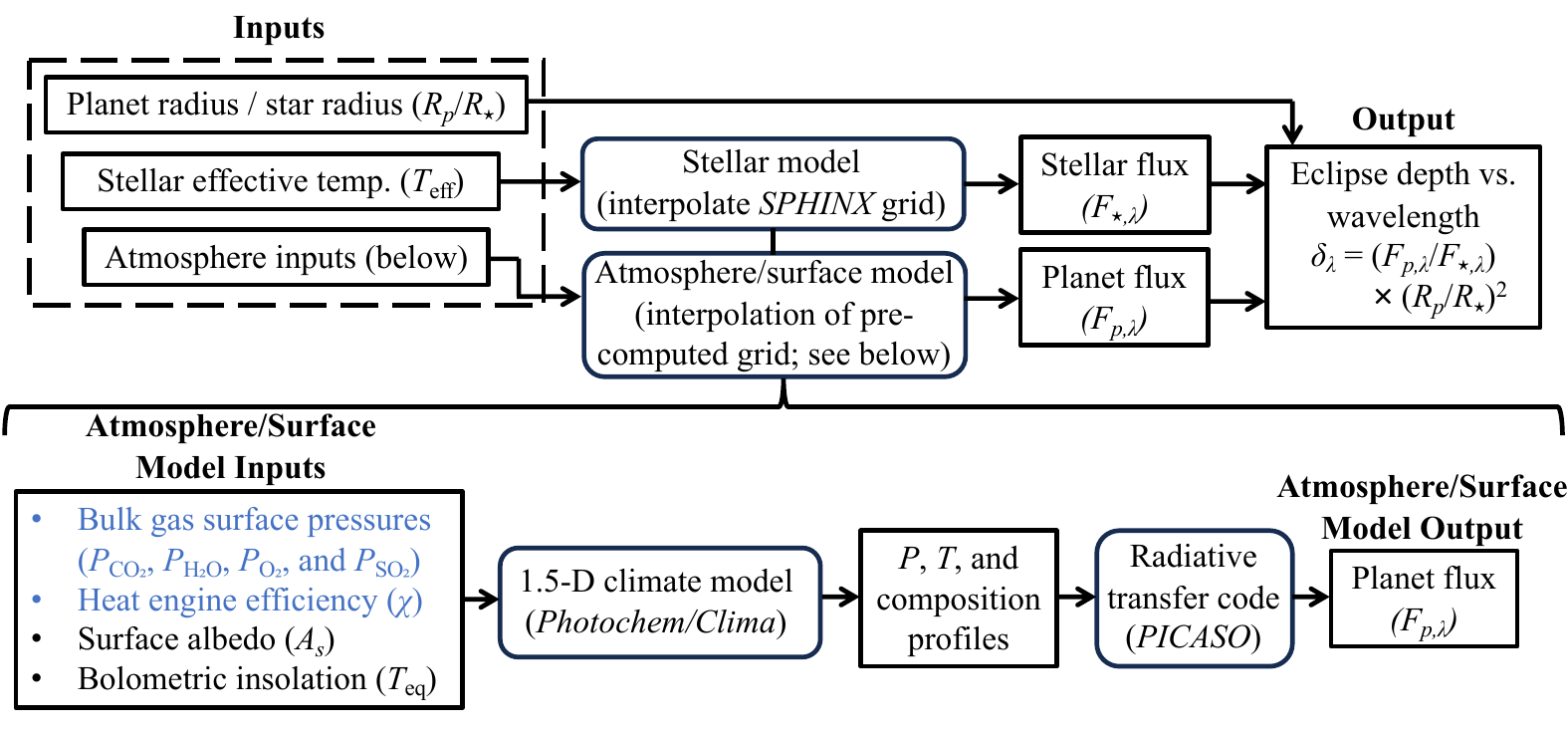}
  \caption{The atmospheric model that we fit, using Bayesian nested sampling, to the MIRI LRS eclipse spectrum of LTT 1445A b. The top half of the figure is the full forward model and the bottom half details the atmospheric and surface component. The output is eclipse depth versus wavelength, which can be compared to MIRI spectra. Inputs (black dashed box) include $R_p/R_\star$, the stellar effective temperature ($T_\mathrm{eff}$), and seven inputs to the atmosphere/surface model. The five atmosphere/surface model inputs in blue text are specific to the atmosphere, and are not considered in our bare-rock fit to the MIRI data (Section \ref{sec:methods_bare}).}
  \label{fig:fm}
\end{figure*}

Figure \ref{fig:fm} shows the atmospheric model that we fit to the JWST eclipse observations. As input, the model takes in nine total parameters, producing as output the eclipse depth as a function of wavelength ($\delta_\lambda$):
\begin{equation} \label{eq:delta}
  \delta_\lambda = \frac{F_{p,\lambda}}{F_{\star,\lambda}} \left(\frac{R_p}{R_\star}\right)^2
\end{equation}
Here, $F_{p,\lambda}$ and $F_{\star,\lambda}$ are the planet and stellar flux as a function of wavelength, respectively. $R_p/R_\star$ is the planet to star radius ratio and is one of the input parameters. 

To compute the stellar flux ($F_{\star,\lambda}$), we linearly interpolate the \texttt{SPHINX} stellar grid \citep{Iyer2023} version 4.0 for a given stellar effective temperature ($T_\mathrm{eff}$). We fix the stellar metallicity and surface gravity to the median values inferred for LTT 1445 A, $\lbrack\mathrm{Fe/H}\rbrack = -0.34$ and $\log_{10}(g) = 4.97$ \citep{Winters2019,Winters2022}, because the measured uncertainties in these quantities have a negligible effect on the predicted eclipse depth. Furthermore, we assume a stellar C/O ratio of 0.5.

The planet flux ($F_{p,\lambda}$) is computed with the atmospheric climate and spectral model described in Section \ref{sec:methods_atmos1}, and illustrated on the bottom-half of Figure \ref{fig:fm}. In brief, given seven input parameters, the atmospheric model computes a temperature profile at radiative-convective equilibrium, then calculates the resulting atmosphere's emission spectra ($F_{p,\lambda}$). The atmospheric code is relatively slow ($\sim10$ core-seconds per model), so we precompute a finely spaced grid then linearly interpolate the spectra during Bayesian inversion (see Section \ref{sec:methods_atmos2} for details). A future update to the framework will consider a full inversion without a precomputed grid, allowing for higher dimensional parameter spaces.

\subsubsection{Climate and spectral model} \label{sec:methods_atmos1}

To compute the planetary flux ($F_{p,\lambda}$), we first simulate an atmosphere's pressure, temperature and composition structure with the 1.5D climate module (\texttt{Clima}) in the \texttt{Photochem} software package version 0.8.2 \citep{Photochem2026}. \citet{Wogan2025b} describes the climate model in full, and validates it against the observed temperature profiles of Venus, Earth, Mars, Jupiter and Titan. Here we provide a brief overview of the code and the updates since the version published in \citet{Wogan2025b}.

\texttt{Clima} solves for radiative and convective equilibrium (RCE) in the atmosphere. To estimate the radiation field, the code uses standard two-stream methods \citep{Toon1989} with opacities representing UV continuum absorption, Rayleigh scattering, collision-induced absorption (CIA) and approximates line absorption with k-distributions \citep{Amundsen2017}. Here, we assume the atmosphere is only composed of \ce{H2O}, \ce{CO2}, \ce{O2} and \ce{SO2} (justified below), and so only include opacities for those species. Most opacities are derived from HITEMP or HITRAN \citep{Gordon2017,Hargreaves2020}, although \citet{Wogan2025b} provides a complete description of each opacity and their origin in the literature. Convection occurs in the model if the temperature gradient exceeds a moist or dry pseudo-adiabat \citep{Graham2021}, although no condensation occurs in our simulations of LTT 1445A b, so only the dry adiabat is relevant. The code estimates day-to-night heat redistribution with a parameterization from \cite{Koll2022}, making the model ``1.5-D''.

In this study, the climate model has seven free input parameters: the surface partial pressures of \ce{CO2}, \ce{H2O}, \ce{O2}, and \ce{SO2} ($P_\mathrm{CO_2}$, $P_\mathrm{H_2O}$, $P_\mathrm{O_2}$, and $P_\mathrm{SO_2}$), the heat-engine efficiency ($\chi$), the surface albedo ($A_s$), and the bolometric insolation, which we parameterize using the zero-albedo equilibrium temperature ($T_{\mathrm{eq},0}$). The gas partial pressures determine both the total surface pressure, $P_s = \sum_i P_i$, and the atmospheric composition. We assume the atmosphere is vertically well-mixed. The heat-engine efficiency, $\chi$, enters the day-night heat redistribution parameterization. \citet{Koll2022} suggest a nominal value of 0.2, but note an uncertainty of at least a factor of two, so we treat $\chi$ as a free parameter. The surface albedo, $A_s$, sets the reflectivity of the surface to incoming stellar radiation and the surface emissivity, $\epsilon = 1 - A_s$, used in the thermal radiative transfer.

Although the eclipse calculation in Equation \ref{eq:delta} allows the stellar flux ($F_{\star,\lambda}$) to vary through $T_\mathrm{eff}$, the climate grid is computed assuming a fixed shape of the stellar spectrum (although with variable bolometric flux at the planet). Specifically, each climate simulation adopts a single \texttt{SPHINX} stellar spectrum with $T_\mathrm{eff} = 3340$ K \citep{Winters2022}, because we find that the measured uncertainty in $T_\mathrm{eff}$ \citep[$\pm150$ K;][]{Winters2019} has a negligible effect on the pressure-temperature profile and planetary emission spectrum. Likewise, for all climate simulations (but not in Equation \eqref{eq:delta}) we fix the planet radius and mass to 1.34 $R_{\oplus}$ and 2.73 $M_{\oplus}$ \citep{Pass2023}, respectively, because the measured uncertainties in these quantities also have a small effect on the P-T profile and planet emission.

We restrict the atmospheric composition to \ce{CO2}, \ce{H2O}, \ce{O2}, and \ce{SO2} for three reasons. First, LTT 1445A b likely experienced substantial XUV irradiation during its host star's extended pre-main-sequence phase, which would have driven efficient escape of hydrogen and helium. Consequently, if the planet retains an atmosphere today, we expect H-bearing species such as \ce{H2}, \ce{CH4}, and \ce{NH3} to be disfavored. Consistent with this picture, \citet{KrissansenTotton2024} found that erosion of a primary atmosphere from a warm rocky planet orbiting an M star commonly yields a tenuous \ce{O2}- or \ce{CO2}-rich atmosphere, and other evolutionary studies have reached similar conclusions \citep[e.g.,][]{Luger2015,Cherubim2025}. Second, we do not include plausible, but radiatively weak gases like \ce{CO} and \ce{N2} because we have found they have a similar impact on the P-T profile and emission spectrum as \ce{O2}, as shown in Appendix Figure \ref{fig:surf_thin} (right panel). In other words, the input \ce{O2} pressure can be thought of as the sum of \ce{O2}, \ce{CO}, \ce{N2} and any other feasible but low opacity species. Finally, we omit \ce{O3}, a photochemical bi-product of \ce{O2}. Appendix \ref{sec:appendix_photochemistry} uses a self-consistent coupled climate-photochemical model to show that the predicted \ce{O3} does not change the temperature or the emission spectrum (near the $\sim 9$ $\mu$m \ce{O3} feature) by an amount that is relevant to the existing low-precision MIRI LRS data. 

The pressure-temperature-composition profiles generated by the climate code are passed to \texttt{PICASO} \citep{Batalha2019} to compute the planetary emission spectrum. We use $R = \mathrm{15,000}$ resampled opacities archived on Zenodo \citep{Opacities2025}, derived from the same set of HITEMP and HITRAN opacities used by \texttt{Photochem}/\texttt{Clima}. In brief, we account for line absorption from \ce{CO2}, \ce{H2O}, \ce{O2}, and \ce{SO2} as well as the collision-induced absorption (CIA) partners \ce{H2O}-\ce{H2O}, \ce{O2}-\ce{O2}, \ce{CO2}-\ce{CO2}. The \ce{H2O}-\ce{H2O} absorption is a reformulation of the MT\_CKD water continuum \citep{Mlawer2012} as a CIA opacity. Our primary analysis assumes clear-sky atmospheres, but Appendix Section \ref{sec:appendix_clouds} considers the impact of aerosols. We down bin to a resolution of 100, before saving the spectra in our grid of calculations (Section \ref{sec:methods_atmos2}).

\subsubsection{Precomputed grid and interpolation} \label{sec:methods_atmos2}

\begin{table*}
  \centering
  \caption{Precomputed atmospheric grid and adopted priors used in the Bayesian inversion.}
  \label{tab:grid_priors}
  \begin{tabularx}{0.65\textwidth}{p{0.1\textwidth} p{0.2\textwidth} p{0.3\textwidth}}
    \hline
    Parameter & Grid values & Prior \\
    \hline
    $\log_{10}(P_{\mathrm{H_2O}})$ & $-7$ to $2$ in steps of $1$ & $\mathcal{U}(-7, 2)$ \\
    $\log_{10}(P_{\mathrm{CO_2}})$ & $-7$ to $2$ in steps of $0.5$ & $\mathcal{U}(-7, 2)$ \\
    $\log_{10}(P_{\mathrm{O_2}})$ & $-5$ to $2$ in steps of $1$ & $\mathcal{U}(-5, 2)$ \\
    $\log_{10}(P_{\mathrm{SO_2}})$ & $-7$ to $2$ in steps of $1$ & $\mathcal{U}(-7, 2)$ \\
    $\chi$ & $0.05$, $0.2$, and $0.8$ & $\mathcal{U}(\log_{10}(0.05), \log_{10}(0.8))$ \\
    $A_s$ & 0.0, 0.1, 0.2, 0.3 and 0.4 & $\mathcal{U}(0, 0.4)$ \\
    $T_{\mathrm{eq},0}$ & $385$, $431$, and $477$ K & $\mathcal{TN}(431, 23; 385, 477)$ \\
    $T_\mathrm{eff}$ & N/A & $\mathcal{TN}(3340, 150; 3040, 3640)$ \\
    $R_p/R_\star$ & N/A & $\mathcal{TN}(0.0454, 0.0012; 0.0430, 0.0478)$ \\
    \hline
    \multicolumn{3}{p{0.65\textwidth}}{
      Note: $\log_{10}(P_i)$ is the surface partial pressure of species $i$ in $\log_{10}$ bars. $\mathcal{U}(a,b)$ denotes a uniform prior on the interval $[a,b]$, and $\mathcal{TN}(\mu,\sigma;a,b)$ is a normal prior with mean $\mu$ and standard deviation $\sigma$, truncated to the interval $[a,b]$.
    }\\
  \end{tabularx}
\end{table*}

As mentioned previously, the climate and spectral model described in Section \ref{sec:methods_atmos1} is slow enough that we choose to precompute a large grid of atmospheres and spectra, which we then linearly interpolate during a Bayesian inversion. The first two columns of Table \ref{tab:grid_priors} summarize the grid, which totals 684,000 model runs. The full calculation took $\sim 3.5$ hours using 512 CPU cores on the NASA Aitken Supercomputer (4 nodes each with 128 cores). We justify the ranges for each parameter when detailing our adopted priors in Section \ref{sec:methods_inversion}. To verify that linear interpolation of the precomputed grid is sufficiently accurate, we compared 100 interpolated spectra against spectra computed directly with the full climate and spectral model at randomly sampled parameter values between grid points. Across all tests, the median and maximum error in eclipse depth at the data's resolution was 0.6 and 6.9 ppm, respectively, both values smaller than the $\sim20$ ppm 1$\sigma$ error reported for MIRI LRS data binned to 16 points. Furthermore, the largest interpolation errors occurred between $\sim 8$ to 11 $\mu$m in the LRS bandpass, where the data has the worst precision (between 20 and 35 ppm).

\subsection{Bare-rock model} \label{sec:methods_bare}

We also fit a bare-rock model to the MIRI LRS spectrum that is nested within the full atmospheric model. The bare-rock model has four free parameters: the zero-albedo equilibrium temperature ($T_{\mathrm{eq},0}$), the surface albedo ($A_s$), the stellar effective temperature ($T_\mathrm{eff}$), and the planet-to-star radius ratio ($R_p/R_\star$). For a given input $T_{\mathrm{eq},0}$, we first compute the bolometric flux at the planet, $F_\mathrm{bol} = 4 \sigma T_{\mathrm{eq},0}^4$, where $\sigma$ is the Stefan-Boltzmann constant. The effective dayside temperature is given by the following expression, for a redistribution factor of $f = 2/3$:
\begin{equation}
  T_\mathrm{day} = \left(\frac{f F_\mathrm{bol}}{\sigma}\right)^{1/4}
\end{equation}
The planetary emission is then a blackbody at temperature $T_\mathrm{day}$ with flux
\begin{equation}
  F_{p,\lambda} = \epsilon \pi B_\lambda(T_\mathrm{day}),
\end{equation}
where $B_\lambda$ is the Planck function, and $\epsilon$ is the emissivity of the surface set to $1 - A_s$. As in the atmospheric model, we interpolate the \texttt{SPHINX} grid for the stellar flux and then compute the eclipse depth with Equation \ref{eq:delta}. This bare-rock model is the no-atmosphere limit of the atmospheric model, corresponding to the case in which each gas partial pressure is set to the lowest value in Table \ref{tab:grid_priors}.

\subsection{Bayesian inversion} \label{sec:methods_inversion}

We use the \texttt{PyMultiNest} package \citep{Buchner2014} to fit both the atmospheric and bare-rock models to the eclipse spectrum of LTT 1445A b. The atmospheric inversion considers nine free parameters listed in the first column of Table \ref{tab:grid_priors}, while the bare-rock inversion fits for four parameters ($A_s$, $T_{\mathrm{eq},0}$, $T_\mathrm{eff}$, and $R_p/R_\star$). Each calculation adopts 1000 live points and a standard evidence tolerance of 0.5 as convergence criteria. We nominally use the 16-bin version of the MIRI LRS spectrum published by \citet{Wachiraphan2025}, but also consider the 8-bin dataset as a sensitivity test.

The third column of Table \ref{tab:grid_priors} gives the adopted prior distributions for all inversions. Gas partial pressures have log-uniform priors between $10^{-7}$ or $10^{-5}$ bar and $10^{2}$ bar. The thinnest atmosphere in this prior space have emission spectra very similar to a bare rock. The heat-engine efficiency ($\chi$) has a log-uniform prior between 0.05 and 0.8, spanning a factor of 4 below and above the recommended value of 0.2 \citep{Koll2022}. Surface albedo has a uniform prior between 0 and 0.4. As shown in Appendix Figure \ref{fig:surf_thin} (left panel), this range approximately encompasses the expected emission from low-albedo surfaces like ``basaltic'' and high-albedo surfaces like ``granitoid'' \citep{Mansfield2019}. Planets commonly interpreted as bare rocks, such as TRAPPIST-1 b, appear to have dark surfaces with albedos near zero \citep{Greene2023}, so allowing values as high as $A_s = 0.4$ is conservative, and follows from the recommendations of \citet{Mansfield2019}. The priors on $T_{\mathrm{eq},0}$, $T_\mathrm{eff}$, and $R_p/R_\star$ are truncated normal distributions with means and standard deviations taken from \citet{Winters2022} and \citet{Pass2023}, truncated at $\pm 2 \sigma$ from the mean.

\section{Results} \label{sec:results}

\subsection{Primary Atmospheric Constraints}

\begin{figure*}
  \centering
  \includegraphics[width=.9\textwidth]{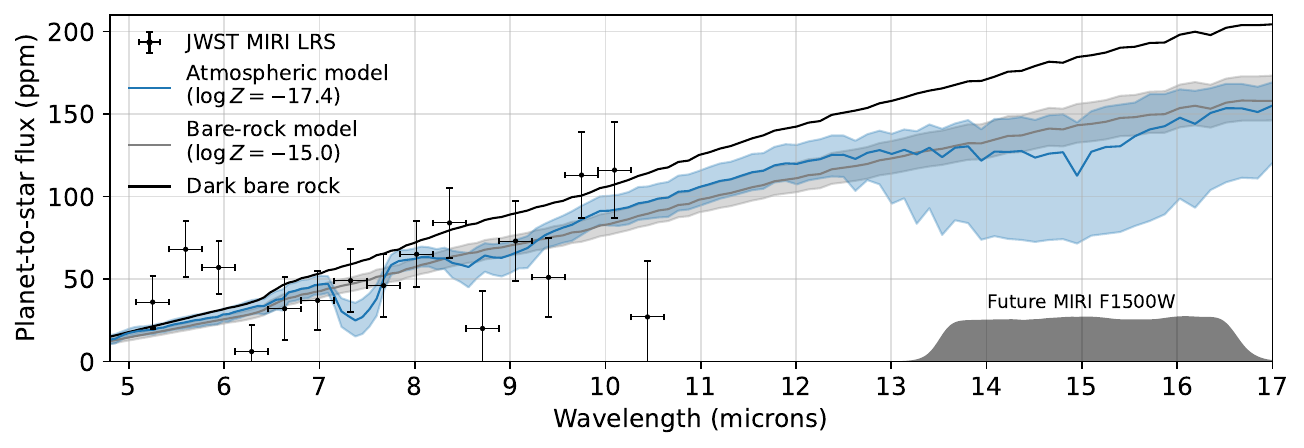}
  \caption{The atmospheric and bare-rock fits to the 16-bin MIRI LRS eclipse spectrum. Lines and shaded bands show the median model spectrum and 1$\sigma$ credible interval. The black line shows the expected eclipse depth for a dark bare rock, assuming the mean planetary and stellar parameters of LTT 1445A b. Gray shading near 15 $\mu$m marks the throughput curve of the planned MIRI F1500W photometric observations. The bare-rock model is moderately favored by the Bayesian evidence, with $\log Z_\mathrm{bare} = -15.0$ and $\log Z_\mathrm{atm} = -17.4$.}
  \label{fig:spectra}
\end{figure*}



\begin{table*}
  \centering
  \caption{Summary of the Bayesian inversions for the 16-bin and 8-bin MIRI LRS datasets.}
  \label{tab:results}
  \begin{tabularx}{0.63\textwidth}{l c c c c}
    \hline
    & \multicolumn{2}{c}{16-bin dataset} & \multicolumn{2}{c}{8-bin dataset} \\
    Parameter & Atmosphere & Bare rock & Atmosphere & Bare rock \\
    \hline
    $k$ & 9 & 4 & 9 & 4 \\
    $\log Z$ & $-17.35$ & $-14.97$ & $-7.39$ & $-5.44$ \\
    AIC & $43.8$ & $34.7$ & $23.8$ & $16.2$ \\
    $\chi^2_\mathrm{min}$ & $25.8$ & $26.7$ & $5.8$ & $8.2$ \\
    \hline
    $K_{\mathrm{bare/atm}}$ & \multicolumn{2}{c}{$10.8$} & \multicolumn{2}{c}{$7.0$} \\
    $\Delta \mathrm{AIC}_{\mathrm{bare-atm}}$ & \multicolumn{2}{c}{$-9.1$} & \multicolumn{2}{c}{$-7.6$} \\
    \hline
    $\log_{10}(P_\mathrm{s}/\mathrm{bar})$ & $< -0.21$ & --- & $< 0.26$ & --- \\
    $\log_{10}(P_{\mathrm{H_2O}}/\mathrm{bar})$ & $< -2.61$ & --- & $< -2.17$ & --- \\
    $\log_{10}(P_{\mathrm{CO_2}}/\mathrm{bar})$ & $< -1.00$ & --- & $< -0.63$ & --- \\
    $\log_{10}(P_{\mathrm{O_2}}/\mathrm{bar})$ & $< -0.23$ & --- & $< 0.22$ & --- \\
    $\log_{10}(P_{\mathrm{SO_2}}/\mathrm{bar})$ & $< -4.03$ & --- & $< -3.60$ & --- \\
    $\log_{10}(\chi)$ & $-0.74^{+0.39}_{-0.35}$ & --- & $-0.73^{+0.40}_{-0.36}$ & --- \\
    $A_s$ & $0.22^{+0.10}_{-0.12}$ & $0.23^{+0.11}_{-0.12}$ & $0.24^{+0.09}_{-0.12}$ & $0.25^{+0.10}_{-0.13}$ \\
    $T_{\mathrm{eq},0}$ (K) & $437^{+15}_{-17}$ & $437^{+18}_{-18}$ & $436^{+15}_{-17}$ & $433^{+18}_{-17}$ \\
    $T_\mathrm{eff}$ (K) & $3345^{+117}_{-123}$ & $3338^{+132}_{-136}$ & $3351^{+115}_{-127}$ & $3348^{+133}_{-143}$ \\
    $R_p/R_\star$ & $0.0454^{+0.0010}_{-0.0010}$ & $0.0453^{+0.0011}_{-0.0011}$ & $0.0453^{+0.0010}_{-0.0009}$ & $0.0453^{+0.0011}_{-0.0011}$ \\
    \hline
    \multicolumn{5}{p{0.63\textwidth}}{Note: $k$ is the number of free model parameters. Central values and uncertainties are quoted as the median and 16th/84th percentiles. Upper limits are 2$\sigma$ (97.5th percentile). Bayes factors are computed from the evidences as $K_{\mathrm{bare/atm}} = \exp(\log Z_\mathrm{bare} - \log Z_\mathrm{atm})$. The AIC difference is defined as $\Delta \mathrm{AIC}_{\mathrm{bare-atm}} = \mathrm{AIC}_\mathrm{bare} - \mathrm{AIC}_\mathrm{atm}$, so negative values favor the bare-rock model.}\\
  \end{tabularx}
\end{table*}

Figure \ref{fig:spectra} shows both the atmospheric and bare-rock fits to the 16-bin variant of the MIRI LRS eclipse spectrum. Both inversions provide comparable agreement with the data, and their predicted eclipse spectra differ from each other by no more than about $1\sigma$ at most wavelengths. The atmospheric model does not reproduce the apparent structure in the LRS spectrum between 5 and 6.5 $\mu$m and between 8 and 10.5 $\mu$m, suggesting that these fluctuations could be due to noise rather than atmospheric absorption.

\begin{figure*}
  \centering
  \includegraphics[width=\textwidth]{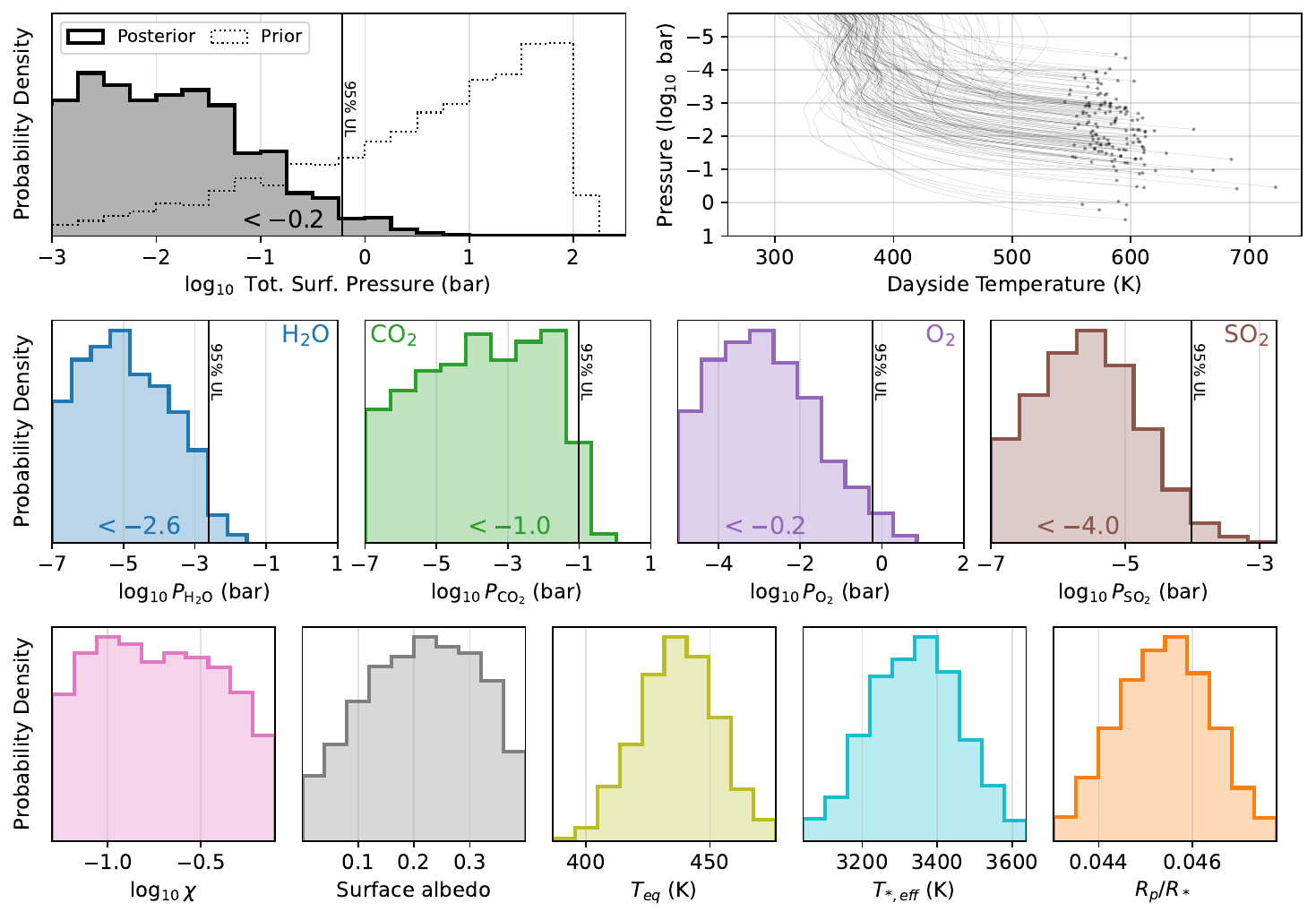}
  \caption{1D marginalized posterior distributions for key parameters in the atmospheric fit to the 16-bin MIRI LRS eclipse spectrum. The top-left panel shows the derived posterior on total surface pressure (solid black line), compared to implicit prior (dotted black line). The top-right panel shows pressure-temperature profiles sampled from the posterior. The climate-constrained inversion places a 95\% upper limit of 0.6 bar on the surface pressure. If an atmosphere is present, a tenuous \ce{CO2}- or \ce{O2}-dominated atmosphere is most probable.}
  \label{fig:retrieval}
\end{figure*}

For the 16-bin MIRI LRS dataset, the atmospheric inversion has a Bayesian evidence of $\log Z_\mathrm{atm} = -17.4$, while the bare-rock fit has the larger value $\log Z_\mathrm{bare} = -15.0$ (Table \ref{tab:results}). From these evidences, we compute a Bayes factor in favor of the bare-rock model of $K_{\mathrm{bare/atm}} = \exp(\log Z_\mathrm{bare} - \log Z_\mathrm{atm}) = 11.0$. Repeating both the atmospheric and bare-rock fits to the 8-bin MIRI dataset yields a slightly smaller but qualitatively similar value of $K_{\mathrm{bare/atm}} = 7.0$. This result indicates that the additional complexity of the atmospheric model is not justified by the current data, and that a bare-rock interpretation provides a simpler explanation of the observations.

Because Bayes factors depend on the adopted priors, and because strongly motivated priors are often difficult to specify for exoplanet atmosphere models \citep{Thorngren2026}, we also compare the models using the Akaike Information Criterion (AIC), which depends on the maximum likelihood penalized by the number of model parameters. We find $\Delta \mathrm{AIC}_{\mathrm{bare-atm}} = -9.1$ and $-7.6$ for the 16-bin and 8-bin MIRI datasets, respectively (Table \ref{tab:results}). Thus, the AIC provides an independent, frequentist check that the additional flexibility of the atmospheric model is not justified over the simpler bare rock model by the current data \citep{Thorngren2026}.

Figure \ref{fig:retrieval} shows the 1D marginalized posterior distributions for the atmospheric fit to the 16-bin MIRI LRS spectrum. No gas is clearly detected, and instead the analysis yields upper limits on the total surface pressure and the pressure of each constituent. In particular, we place a 2$\sigma$ upper limit of $< 0.6$ bar on the total surface pressure (top left panel; $10^{-0.2} \approx 0.6$). Because we adopt independent log-uniform priors on the partial pressure of each gas, the induced prior on total surface pressure is not log-uniform and places relatively more prior weight at high pressures (dotted line in the top-left panel of Figure \ref{fig:retrieval}). A prior chosen to be uniform in total surface pressure would therefore likely yield an even smaller upper limit than the $<0.6$ bar value reported here. 

If there is a tenuous $\sim 0.6$ bar atmosphere, then the composition is most likely dominated by any optically thin gas like \ce{O2} (or \ce{N2} and \ce{CO}). Up to $\sim 0.1$ bar \ce{CO2} is also permitted by the inversion, while only trace \ce{H2O} and \ce{SO2} are allowed ($\lesssim 10^{-3}$ and $\lesssim 10^{-4}$ bar, respectively). The other 5 parameters ($\chi$, $A_s$, $T_{\mathrm{eq},0}$, $T_\mathrm{eff}$ and $R_p/R_\star$) are only weakly constrained and remain broadly similar to their input priors, although the surface albedo shows a mild preference for values near $\sim0.2$. Applying the same atmospheric fit to the 8-bin MIRI dataset yields qualitatively similar results, but with somewhat weaker upper limits on the gas abundances and total surface pressure ($< 1.8$ bar; Table \ref{tab:results}). 

Taken together, these results indicate that the existing MIRI LRS eclipse spectrum can be explained by a simple bare-rock model, and the fit does not significantly improve when an atmosphere is invoked. If an atmosphere is present, we find that one of the thicker atmospheres still consistent with the spectrum to within $\sim2\sigma$ is roughly $\sim1$ bar of \ce{O2} with $\sim0.1$ bar of \ce{CO2}.

\subsection{Sensitivity Tests}

The surface-pressure limits above depend on our parameterization for day-to-night heat redistribution from \citet{Koll2022}. They showed that this analytic approach can overpredict day-to-night heat redistribution for infrared-optically thin atmospheres with surface pressures $\gtrsim 1$ bar, relative to general circulation models (GCMs). In the nominal inversion, we attempted to account for uncertainty in this parameterization by fitting for the heat-engine efficiency term ($\chi$), but this may not capture the full model error.

As a more direct test of this sensitivity, we repeated the climate-constrained inversion for the 16-bin eclipse spectrum using a precomputed climate and spectral grid that assumed zero day-to-night heat redistribution for all atmospheres, regardless of composition or thickness. In this no-redistribution case, we obtain 2$\sigma$ upper limits of 4 bar on the \ce{O2} pressure and 6 bar on the total surface pressure. Even under this extreme assumption, thicker atmospheres are disfavored because they would produce spectral features inconsistent with the data. This constraint comes from the wavelength-dependent shape of the LRS spectrum, not only from the band-integrated dayside brightness temperature. Thus, if the \citet{Koll2022} parameterization overestimates heat redistribution in this regime, an upper limit of about $\sim 6$ bar on the surface pressure is a more conservative bound.

We also tested whether atmospheric aerosols or photochemical processes could alter our interpretation. Appendix \ref{sec:appendix_clouds} argues that direct condensation clouds (e.g., \ce{H2O} and KCl) and photochemical hazes (e.g., \ce{H2SO4}) are unlikely on LTT 1445A b, and so dust is perhaps the most plausible atmospheric aerosol. We simulate a handful of dusty atmospheres and show that dust would likely lower, rather than raise, the inferred upper limit on surface pressure (Appendix Figure \ref{fig:dust}). In other words, dust is unlikely to make a thick atmosphere look like a thin one in the MIRI LRS bandpass. Lastly, we consider how photochemistry could impact our results. Appendix \ref{sec:appendix_photochemistry} shows that photochemical \ce{O3} produces at most a modest $\sim 9$ $\mu$m feature compared to the current LRS uncertainties (Appendix Figure \ref{fig:ozone}), and so the lack of photochemistry in our main analysis is a defensible assumption. In summary, aerosols/clouds and photochemistry do not appear to weaken the conclusion that LTT 1445A b has at most one to several bars of atmosphere dominated by optically thin gas.

\section{Discussion}

\subsection{Comparison to Prior Work}

Our analysis builds on the initial interpretation of \citet{Wachiraphan2025} by turning the LTT 1445A b LRS spectrum into quantitative constraints on atmospheric pressure and composition. \citet{Wachiraphan2025} showed that the data strongly disfavor simple thick \ce{CO2} atmospheres (i.e. $> 1$ bar), while leaving thinner or other compositions uncertain, arguing that more precise data and detailed modeling are needed. Here, we carry out that broader modeling step with climate-constrained Bayesian inversion. Our results are a more restrictive interpretation of the spectrum. A simple bare-rock model fits the data about as well as a more complex atmospheric model. The thickest atmosphere permitted by the spectrum to $2\sigma$ confidence is $\lesssim 1$ bar, dominated by an optically thin gas like \ce{O2}, \ce{N2} or \ce{CO} with $\lesssim 0.1$ bar \ce{CO2}. In an extreme no-redistribution sensitivity test the upper limit on an optically thin gas rises to $\lesssim 5$ bar.

\subsection{Prospects for Future JWST Observations}

There are many planned observations of LTT 1445A b to further search for an atmosphere. General Observer programs 2512 (PIs: Batalha and Teske), 7251 (PI: Bennett) and 7073 (PIs: Lustig-Yaeger and Stevenson) are in the process of collecting numerous NIRISS, NIRSpec and NIRCam transmission observations of the planet. Also, the JWST Rocky Worlds DDT program will collect MIRI F1500W secondary eclipses of LTT 1445A b, targeting the 15 $\mu$m \ce{CO2} feature. Figure \ref{fig:spectra} shows that the atmospheric fit permits a wide range of eclipse depths near 15 $\mu$m, while the bare-rock fit is much more tightly constrained; future F1500W observations therefore have the potential to distinguish between the atmosphere and bare-rock hypotheses.

Focusing on the new eclipse observations, we ask what F1500W precision would be needed for a joint analysis of the LRS and F1500W observations to strongly favor the presence of an atmosphere. This exercise assumes that LTT 1445A b has one of the thicker atmospheres consistent with the LRS spectrum to within $\sim2\sigma$: 1 bar of \ce{O2}, 0.01 bar \ce{CO2}, and a 0.2 surface albedo. We also adopt the nominal star and planet parameters in Table \ref{tab:grid_priors}. For this setup, our model predicts a 56.5 ppm eclipse depth in the F1500W photometric bin.

We then performed a series of Bayesian fits to the 16-bin MIRI LRS dataset combined with a synthetic 56.5 ppm MIRI F1500W observation with $1\sigma$ errors ranging from 5 ppm to 40 ppm in 5 ppm increments. We fit both the atmospheric model and the bare rock model described in Section \ref{sec:methods}. A F1500W $1\sigma$ error of 20 ppm yields a Bayes Factor of 35 in preference for the atmospheric model, constituting strong evidence for an atmosphere \citep{Thorngren2026}. This precision also yields a constraint on the \ce{CO2} pressure close to the true value: $\log_{10}P_\mathrm{CO_2} = -1.59^{+0.51}_{-0.78}$. Errors higher than 20 ppm (e.g., 25 ppm), do not yield strong atmospheric detections or a peaked \ce{CO2} posterior. This analysis does not fold in random noise (i.e., the F1500W data point is centered on truth), so perhaps a slightly higher precision is justified.

To determine the number of MIRI F1500W eclipses required to achieve 20 ppm precision, we use the JWST ETC \citep{Pontoppidan2016}. Assuming 2 hours of out-of-eclipse observations and the 1.39 hour eclipse duration, the ETC predicts the planet-to-star flux can be measured with 28 ppm error for a single visit. We further inflate this error to 36 ppm, because the F1500W observations of TRAPPIST-1 b had a precision 29\% larger than predicted by the ETC \citep{Greene2023}. If precision scales as $\propto 36~{\rm ppm}/\sqrt{n_\mathrm{visit}}$, then we estimate at least 4 visits will be needed to achieve a $\sim20$ ppm error with F1500W. Thus, $\geq 4$ F1500W eclipses should be sufficient to test one of the thickest \ce{O2}-dominated atmospheres still allowed by the LRS data.




\section{Conclusions}

Many previous analyses of JWST rocky planet eclipse spectra employ a sparse forward-modeling approach to evaluate the data for the presence or absence of an atmosphere. Here, we developed a climate-constrained Bayesian inference framework for these datasets that thoroughly marginalizes over the parameter space and uncertainties, providing quantitative surface pressure and atmospheric constraints. The approach restricts the inference to physically self-consistent P-T profiles that satisfy radiative-convective equilibrium. We applied the method to the MIRI LRS eclipse spectrum of LTT 1445A b published in \cite{Wachiraphan2025}. Our main conclusions are as follows.

\begin{itemize}
  \item The LTT 1445A b LRS spectrum can be explained by a bare rock. Model comparison between a bare-rock and atmospheric fit yields a Bayes factor of $\sim 10$ in preference for the bare-rock model. This result should not be interpreted as strong evidence against any remaining atmosphere; rather, it shows that an atmosphere does not need to be invoked to explain the observations.
  \item Our nominal inversion constrains the total surface pressure to $\lesssim 1$ bar at 2$\sigma$ confidence, with this upper limit set by atmospheres dominated by infrared-optically thin gas. This result depends on our parameterization for day-to-night heat redistribution from \citet{Koll2022}, which may overestimate heat redistribution near our inferred pressure limit. As an extreme test, we performed an inversion assuming no day-to-night heat redistribution, finding that optically thin atmospheres up to several bars remain allowed; larger pressures are disfavored by the wavelength-dependent structure of the LRS spectrum.
  \item If LTT 1445A b has a $\sim 1$ bar atmosphere, then it is dominated by an optically thin gas like \ce{O2}, \ce{N2} or \ce{CO}. The spectrum permits up to $\sim0.1$ bar of \ce{CO2}, while \ce{H2O} and \ce{SO2} are constrained to low abundances ($\lesssim 10^{-3}$ and $\lesssim 10^{-4}$ bar, respectively).
  \item The Rocky Worlds DDT program will collect further secondary eclipse observations of LTT 1445A b with MIRI F1500W. We show that if the program achieves a precision of 20 ppm or better, then a combined analysis of the LRS and F1500W would detect one of the thicker atmospheres permitted by the LRS spectrum: $\sim 1$ bar \ce{O2} and $0.01$ bar \ce{CO2}. We estimate this precision can be attained in $\gtrsim 4$ visits.
\end{itemize}

This application to LTT 1445A b demonstrates that climate-constrained Bayesian inversion is a useful tool for interpreting JWST rocky-planet eclipse observations. The approach can be used on other datasets to derive the quantitative pressure constraints necessary to test population-level hypotheses of atmospheric escape, like the cosmic shoreline.

\begin{acknowledgments}

We thank Joshua Krissansen-Totton for comments on an early draft that improved the manuscript. We also thank Munazza Alam and Johanna Teske for conversations that helped inspire this article. N.W. and N.B. acknowledges partial support from NASA's Interdisciplinary Consortia for Astrobiology Research (grant No. NNH19ZDA001N-ICAR) under award number 19-ICAR19\_2-0041. J.L.Y. and K.B.S. acknowledge funding support from the CHAMPs (Consortium on Habitability and Atmospheres of M-dwarf Planets) team, issued by NASA under Grant No. 80NSSC23K1399 through the Interdisciplinary Consortia for Astrobiology Research (ICAR) program.  J.I. was funded through support for JWST Program GO 3730, provided through a grant from the STScI under NASA contract NAS5-03127. 

\end{acknowledgments}

\appendix
\setcounter{figure}{0}
\setcounter{table}{0}
\renewcommand{\thefigure}{A\arabic{figure}}
\renewcommand{\thetable}{A\arabic{table}}

\section{Forward-models that justify inversion priors and approximations}

\begin{figure*}
  \centering
  \includegraphics[width=\textwidth]{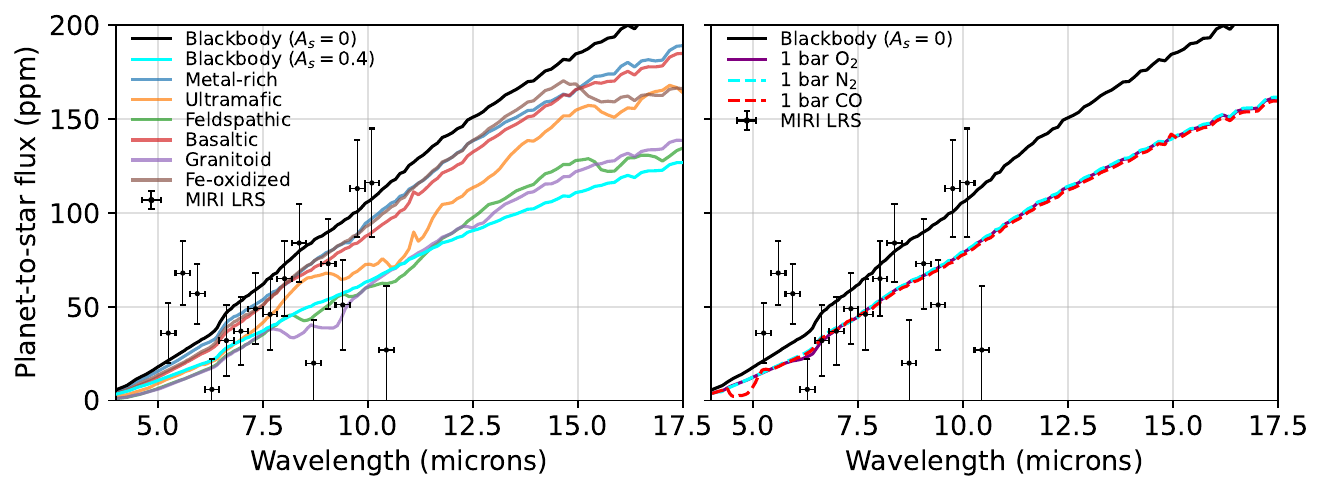}
  \caption{Forward-models used to define the priors for the climate-constrained inversion. Left: bare-rock emission spectra for several plausible surface compositions, compared to blackbody surfaces with $A_s=0$ and $A_s=0.4$. These models motivate our uniform prior on surface albedo between 0 and 0.4. Right: emission spectra for 1 bar atmospheres composed of \ce{O2}, \ce{N2}, or CO, all with surface albedos of 0.2. These simulations have nearly identical emission spectra, motivating our treatment of \ce{O2} as a proxy for other optically thin gases such as \ce{N2} and CO.}
  \label{fig:surf_thin}
\end{figure*}

Appendix Figure \ref{fig:surf_thin} shows supplemental forward-models that motivate choices in our inversion setup. The left panel compares bare-rock emission spectra for several plausible surface compositions from \citet{Hu2012}, supporting the $A_s=0$--0.4 prior range (Section \ref{sec:methods_inversion}). The right panel compares 1 bar \ce{O2}, \ce{N2}, and CO atmospheres, showing that these optically thin gases have nearly identical spectra in the MIRI LRS bandpass, justifying our choice to use \ce{O2} to approximately represent the sum of all low-opacity molecules (Section \ref{sec:methods_atmos1}).

\section{Aerosols and Clouds} \label{sec:appendix_clouds}

\begin{figure}
  \centering
  \includegraphics[width=\linewidth]{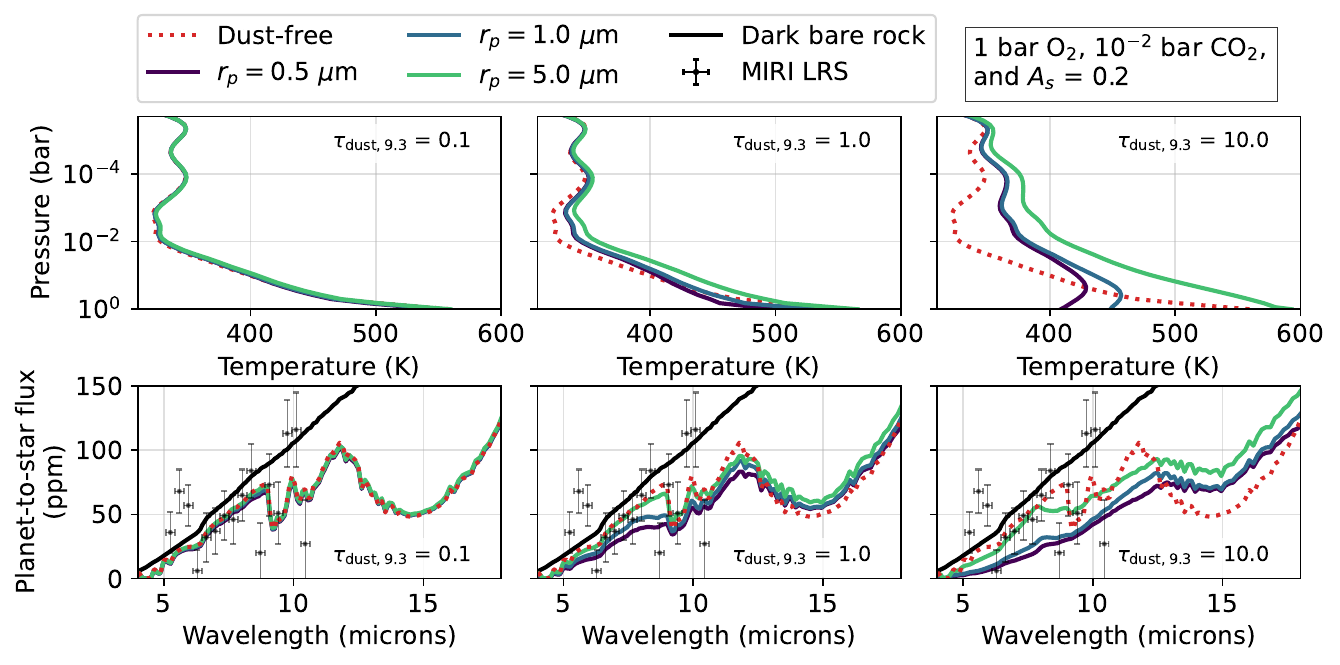}
  \caption{The impact of dust on the P-T profile and emission spectrum of LTT 1445A b. All models have 1 bar of \ce{O2}, 0.01 bar \ce{CO2} and a 0.2 surface albedo. The red dotted line is an aerosol-free model, compared to dusty-simulations with various particle radii and $9.3$ $\mu$m dust column optical depths. Dust mutes spectral features, causing the atmosphere to have lower emission in the LRS bandpass and higher emission near 15 $\mu$m.}
  \label{fig:dust}
\end{figure}

Our climate-constrained inversion does not include atmospheric aerosols or clouds, which can modify the thermal structure and emission spectrum. Here, we first argue that dust is the most plausible atmospheric aerosol for LTT 1445A b. We then show that even if dust were included, it would likely strengthen, rather than weaken, our conclusion that the planet has at most several bars of atmosphere dominated by optically thin gas.

Cloud formation by direct condensation is unlikely. Given LTT 1445A b's bolometric insolation, \ce{H2O} and KCl are two of the most plausible species that could directly condense to clouds \citep{Batalha2026}. To test whether either species is likely to condense, we evaluated high-likelihood posterior atmospheres from the 16-bin inversion, defined as models within 4.5 $\Delta \log L$ of the maximum-likelihood solution, roughly equivalent to a $3\sigma$ interval for one Gaussian parameter. \ce{H2O} was undersaturated by a factor of $\sim 3500$ across all the samples. The maximum KCl saturation surface mixing ratio is only $1.6\times10^{-7}$ by volume, implying that little KCl vapor could be available to form optically significant clouds.

It is also challenging to make a case for photochemically produced aerosols. Section \ref{sec:methods_atmos1} argues that significant hydrogen-bearing species are not expected on LTT 1445A b given its XUV irradiation history, meaning we should have a low prior for hydrocarbon hazes derived from \ce{CH4} photochemistry. Sulfur-based photochemical hazes composed of \ce{H2SO4} and \ce{S8} are also unlikely. Across the high likelihood ($3\sigma$) posterior samples described in the previous paragraph, the minimum saturation mixing ratio for \ce{H2SO4} and \ce{S8} are far above 1, indicating that neither species can condense anywhere in the modeled atmospheric columns.

By process of elimination, we argue the most likely atmospheric aerosol on LTT 1445A b is dust lofted from the surface, much like the dust suspended in Mars' atmosphere \citep{Kahre2023,Montabone2015,Wolff2009}. To assess how dust could impact our interpretation of the LRS data, we simulate several dusty atmospheres self-consistently with our climate code. We assume that dust has a vertically constant particle mixing ratio (i.e., [particles/cm$^2$]/[gas molecules/cm$^2$]), with a magnitude parameterized by the $9.3$ $\mu$m dust column optical depth. This simple parameterization is motivated by Mars, where dustiness is often quantified by the $9.3$ $\mu$m column optical depth \citep[i.e.,$\tau_{\mathrm{dust},9.3}$;][]{Montabone2015,Wogan2025}. For context, Mars has $\tau_{\mathrm{dust},9.3}\sim0.1$--0.15 during relatively clear seasons, while values of order $\gtrsim 1$ correspond to severe dust-storm conditions \citep{Montabone2015}. Each calculation assumes an atmosphere with 1 bar \ce{O2}, 0.01 bar \ce{CO2} and $A_s = 0.2$, and adopts Mie optical properties for dust from \citet{Wolff2009} based on Mars observations. 

Figure \ref{fig:dust} shows the clear-sky simulation (red dotted lines), compared to dusty atmospheres with various particle radii ($r_p$) and dust loadings, quantified with $\tau_{\mathrm{dust},9.3}$. Larger values of $\tau_{\mathrm{dust},9.3}$ with small particle sizes lead to warmer upper atmospheres from starlight absorption by dust. Relative to the clear-sky emission spectrum, dusty atmospheres mute molecular spectral structure and lower the planet-to-star flux across much of the LRS bandpass, but they can increase the flux near $15$ $\mu$m in the MIRI F1500W bandpass. In this sense, dust makes a given atmosphere appear thicker in the LRS data. As a result, including dust in the climate-constrained inversion would likely lower, rather than raise, the upper limit on surface pressure.

\section{Photochemistry} \label{sec:appendix_photochemistry}

\begin{figure}
  \centering
  \includegraphics[width=\linewidth]{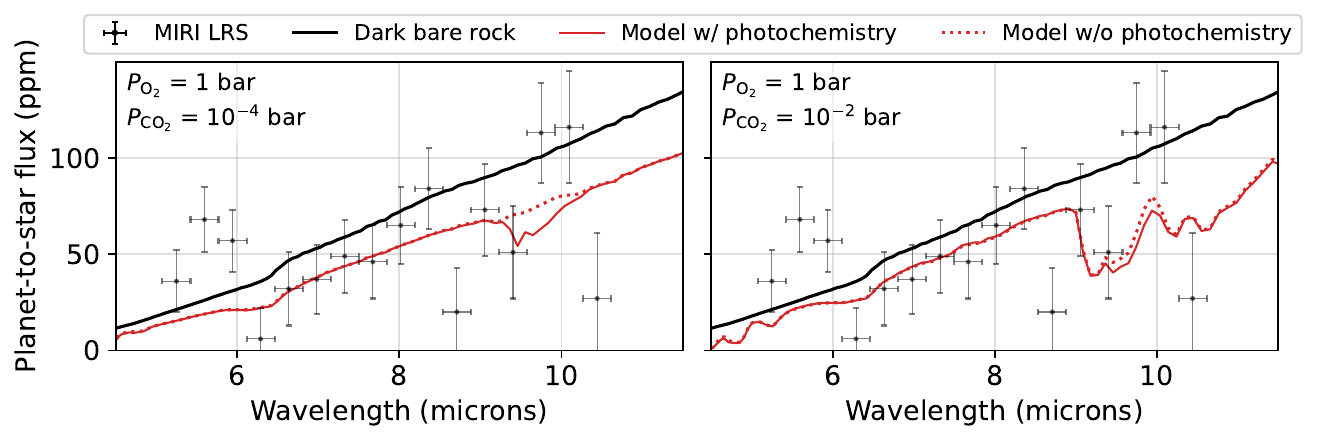}
  \caption{The impact of photochemistry on the emission spectrum of LTT 1445A b. Two atmospheres are simulated: 1 bar \ce{O2} with $10^{-4}$ bar \ce{CO2} (left panel), and 1 bar \ce{O2} with $10^{-2}$ bar \ce{CO2} (right panel). Solid lines are the results of coupled photochemical-climate simulations, and dotted lines are climate-only models. This sensitivity test demonstrates that photochemically produced \ce{O3} has a relatively small impact on the planet's emission spectrum when compared to the existing low-precision data.}
  \label{fig:ozone}
\end{figure}

Photochemistry in a \ce{O2}-rich atmosphere can generate \ce{O3}, which absorbs within the LRS bandpass near $\sim 9,\mu$m. Our climate-constrained inversion does not include photochemistry, and so does not account for the possible impact of \ce{O3}. To test whether \ce{O3} could produce a feature large enough to alter our interpretation, we performed simulations of LTT 1445A b that couple the photochemical and climate models in \texttt{Photochem}. Specifically, we iterated between the two modules until reaching a self-consistent climate and photochemical state. For the photochemical simulation, we used the code's nominal chemical network and adopted the UV spectrum of GJ176 from MUSCLES \citep{France2016} as a proxy for the host star's UV emission. We considered two atmospheres permitted by the climate-constrained retrieval: 1 bar \ce{O2} with $10^{-4}$ bar \ce{CO2}, and 1 bar \ce{O2} with $10^{-2}$ bar \ce{CO2}.

Figure \ref{fig:ozone} shows the emission spectra for the two atmospheres. Solid lines include photochemistry, while dotted lines show the climate-only models. \ce{O3} reaches a peak mixing ratio of about $10^{-5}$ in both cases. The $10^{-4}$ bar \ce{CO2} case produces a $\sim 9,\mu$m \ce{O3} feature with an amplitude of only 16 ppm, smaller than the median 19.5 ppm $1\sigma$ uncertainty in the LRS data. In the $10^{-2}$ bar \ce{CO2} case, a distinct \ce{O3} feature is not apparent because \ce{CO2} also absorbs near $\sim 9,\mu$m. Overall, this sensitivity test shows that photochemical \ce{O3} is unlikely to alter LTT 1445A b's emission spectrum in a way that changes our interpretation of the current low-precision LRS dataset.

\bibliography{bib}
\bibliographystyle{aasjournalv7}

\end{document}